# Microwave Nanotube Transistor Operation at High Bias

Z. Yu, C. Rutherglen, P.J. Burke

*Integrated Nanosystems Research Facility, Department of Electrical Engineering and Computer Science,
University of California, Irvine, CA 92697-2625*

We measure the small signal, 1 GHz source-drain dynamical conductance of a back-gated single-walled carbon nanotube field effect transistor at both low and high dc bias voltages. At all bias voltages, the intrinsic device dynamical conductance at 1 GHz is identical to the low frequency dynamical conductance, consistent with the prediction of a cutoff frequency much higher than 1 GHz. This work represents a significant step towards a full characterization of a nanotube transistor for RF and microwave amplifiers.

Theoretically, single-walled carbon nanotube field effect transistors (SWNT FETs) are predicted to have intrinsic cut-off frequencies approaching the THz range[1, 2]. In this Letter, we take a significant step towards demonstrating this goal by measuring the high dc bias small signal 1 GHz dynamical source-drain conductance of a back-gated SWNT FET. The intrinsic device dynamical conductance at 1 GHz is identical to the low frequency dynamical conductance, consistent with the prediction of a cutoff frequency much higher than 1 GHz.

A full RF characterization of any three terminal device (including SWNT FETs) requires the measurement of a matrix that relates the source ac voltage, the source ac current, the gate ac voltage, and the gate ac current[3]. This 2x2 matrix (called the h-matrix, or equivalently the impedance matrix, or equivalently the S-matrix) depends on the dc voltages at the source and gate, and on the frequency. For typical operating conditions in applications such as low-noise amplifiers (LNAs), one is most interested in the value of this matrix at a dc bias voltage in the range of saturation. This matrix determines the gain of the device in a practical circuit, as well as the input and output impedance, which determines what sort of external impedance matching circuits are needed on the input and the output.

We recently measured[4] one element of this matrix at 2.6 GHz for a SWNT FET at cryogenic temperatures in the zero-dc source-drain bias range of operation. While our work was the first demonstration of SWNT FET operation at microwave frequencies, the zero-bias range of operation is not the most technologically relevant bias, especially for LNAs. In this Letter, we present the first microwave measurements at the technologically relevant high dc-bias voltage of the ratio of the ac source-drain current to the ac source-drain voltage, one element of the y-matix. (This can be considered the drain resistance in our recent model[1].) Using a technique to de-embed parasitics[5], we show that this ratio at 1 GHz is the same as it is at dc for the intrinsic device. This is the first measurement of any of the elements of the S-matrix at high bias for a SWNT FET, and a significant step towards a full characterization of the S-matrix that sets the output impedance, input impedance, and ultimately the gain of a SWNT FET.

Individual SWNTs were synthesized via chemical vapor deposition according to previously published recipes[6, 7] on oxidized, p-doped Si wafers with a 300-400 nm SiO$_2$ layer. Metal electrodes were formed on the SWNTs using electron-beam lithography and metal evaporation of 30-nm Pd/20-nm Cr/100 nm Au trilayer[8]. The devices were not annealed. Nanotubes with electrode spacing of 1 μm were studied. Typical resistances were ~ 100 kΩ; some nanotubes had resistances below 50 kΩ. In this study we focus on semiconducting SWNTs (defined by a gate response) with resistance below 50 kΩ. Measurements were performed at room temperature in air. We fabricated devices on two wafers. One of the wafers contains 32 devices, a yield of 70%, and an average resistance of 140 kΩ. The other wafer contains 30 devices, a yield of 60%, and an average resistance of 130 kΩ.   Fig 1. shows an SEM image of the SWNT device presented in this Letter; other devices showed similar behavior.

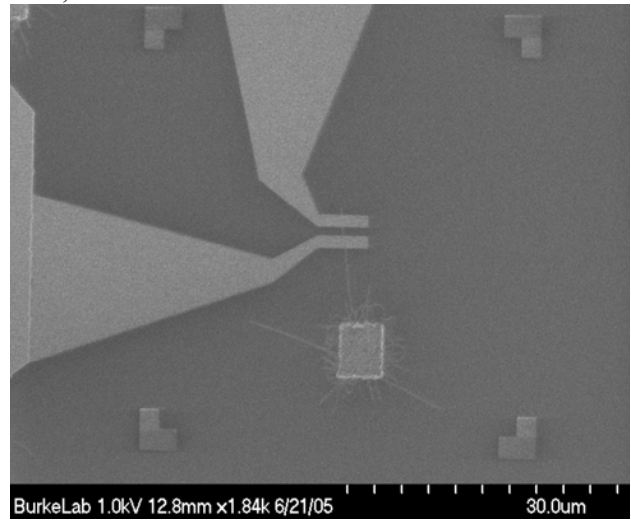

Figure 1: SEM image of SWNT FET.

Fig. 2 shows the room temperature I-V characteristic of a SWNT FET with a 1 μm source-drain spacing. Since this length is comparable to the mean-free-path, this device is in the quasi-ballistic limit at low bias, but in the diffusive regime at high bias. The low-bias resistance of this device was 47 kΩ. The inset shows the low-bias depletion curve, which displayed p-type behavior.

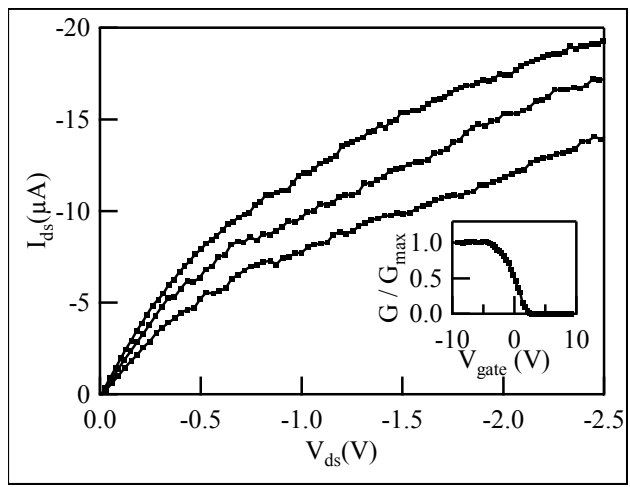

Figure 2: Current-voltage characteristic at Vg=-8, -7, -5 V for SWNT FET. Inset shows small bias depletion curve.

In order to measure the dynamical impedance at microwave frequencies, we employed the same technique that we recently used to measure metallic SWNTs[5]. A commercially available microwave probe (suitable for calibration with a commercially available open/ short/load calibration standard) allowed for transition from coax to lithographically fabricated on chip electrodes. The electrode geometry consisted of one small contact pad of 50x50 $\mu m^2$, and the other 200x200 $\mu m^2$ A microwave network analyzer is used to measure the calibrated (complex) reflection coefficient $S_{11}(\omega) \equiv V_{reflected}/V_{incident}$, where $V_{incident}$ is the amplitude of the incident microwave signal on the coax, and similarly for $V_{reflected}$. This is related to the load impedance $Z(\omega)$ by the usual reflection formula: : $S_{11}=[Z(\omega)-50 \Omega]/[Z(\omega)+50 \Omega]$. At the power levels used (3 $\mu W$), the results are independent of the power used.

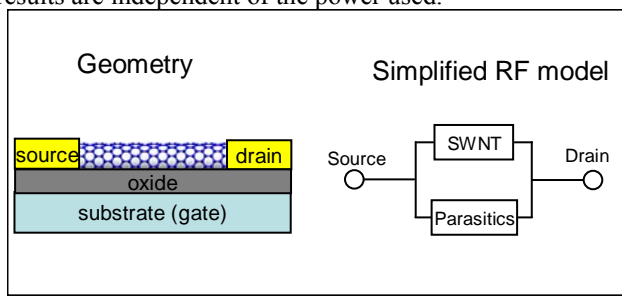

Figure 3: Device geometry and RF model for interpretation.

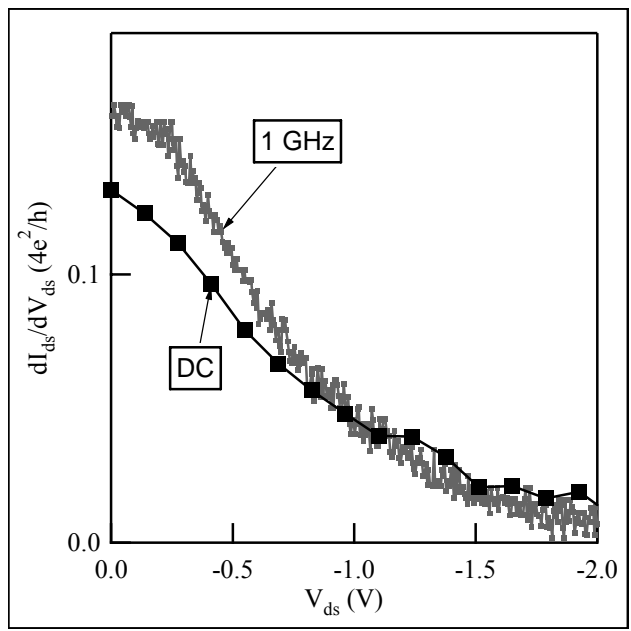

Figure 4: Dynamical conductance at DC and 1 GHz at $V_{gate} = -8$ V.

As we discussed in quantitative detail in our work on metallic SWNTs[5], measurements of the absolute value of the microwave conductance of a high impedance device are generally associated with significant error bars. This is because of the difficulty in separating the inherent device performance from parasitics in parallel with the device, as shown in Fig. 3. Unless extreme care is taken in the calibration, these parasitics are difficult to model and hence "calibrate out". However, *changes* in the device ac properties with dc bias are much clearer since the dc bias can be changed without the need to physically adjust the probes. This assumes that the parasitics do not change with dc bias, which can easily be checked in a control experiment. In our experience, such control experiments are critical since in certain cases the doped semiconducting substrate can change its rf feedthrough properties depending on the electrode dc bias, which can easily be mistaken for intrinsic nanotube performance.

We plot in Fig. 4 the SWNT FET dynamical conductance as a function of source-drain bias voltage for a fixed gate voltage of -8 V. The DC curve is obtained by numerically differentiating the measured I-V curve. The AC (1 GHz) curve is obtained by relating the change in the measured value of $S_{11}$ to the change in the conductance through the formula G(mS) ≈ 1.1 x $S_{11}$(dB)[5]. A background parasitic conductance of 3.37 x $4e^2/h$ was subtracted from G, which was determined using a control set of electrodes integrated onto the same wafer with no nanotube.

The changes in $S_{11}$ with the source-drain voltage are systematic and reproducible. The change in $S_{11}$ with source-drain voltage is not an artifact, since control samples without CNT bridging do not exhibit this effect. As such, this is the first measurement of the dynamical impedance of a SWNT FET at high bias. Our results clearly demonstrate that the 1 GHz intrinsic device performance (neglecting parasitics) is the same as the dc performance, indicating that the ultimate intrinsic speed limit is much larger than our measurement frequency of 1 GHz.

Although prior to this work SWNT FET linear response has only been observed at MHz frequencies[9], recent work has demonstrated *rectification* in SWNT FETs at frequencies up to 50 GHz[10-12]. This interesting and significant work provides quantitative information about device performance as a non-linear rectifier (i.e. diode), and even gives information about the frequency limit of this non-linearity. As nano-devices are predicted to be inherently fast[1], it is a significant achievement to demonstrate the frequency dependence of this non-linearity in order to experimentally prove the ultimate speed limits of nano-devices. However, rectification does not explicitly measure the small-signal S-matrix, especially its frequency dependence, and hence is a significant but somewhat indirect step towards ultrafast amplification of RF signals by a nano-device. In addition, studies to date have focused on large-signal RF voltages, so that the device is operating far out of the linear range. For applications such as amplifiers (both LNAs and power amplifiers), one is interested in *avoiding* non-linearities. In particular, rectification does not address the frequency limitation of the gain of the device, nor impedance matching requirements at the input and output. Both of these requirements are critical challenges that must be addressed head-on before any practical use is to be made of nano-devices of any kind in the RF and microwave frequency range.

For this reason, the measurement of the high bias dynamical source-drain conductance in a SWNT FET presented herein is significant step in that direction. Only when the entire 2x2 matrix has been measured will an RF characterization of SWNT FETs be complete.

Acknowledgements: This work was supported by the Army Research Office (award DAAD19-02-1-0387) and the National Science Foundation (awards ECS-0300557, CCF-0403582, DMR- 0216635).